\documentclass[final,4p]{elsarticle}
\usepackage{graphicx,amsmath,epsfig,amssymb}
\usepackage{txfonts}

\begin{document}

\begin{frontmatter}

\title{Chemical evolution during the process of proto-star formation by considering 
a two dimensional hydrodynamic model}
\author[ICSP]{Ankan Das}
\ead{ankan@csp.res.in}
\author[ICSP]{Liton Majumdar}
\ead{liton@scp.res.in}
\author[SNBNCBS,ICSP]{Sandip K. Chakrabarti}
\ead{chakraba@bose.res.in}
\author[MMCC,ICSP]{Sonali Chakrabarti}
\ead{sonali@csp.res.in}
\address[ICSP]{Indian Centre For Space Physics, 43 Chalantika, Garia Station Road, Kolkata 700084, India}
\address[SNBNCBS]{S.N. Bose National Center for Basic Sciences, JD-Block, Salt Lake, Kolkata,700098, India}
\address[MMCC]{Maharaja Manindra Chandra College, 20 Ramakanto Bose Street, Kolkata, 700003,India}

%\date{Accepted; Received }

\begin{abstract}
Chemical composition of a molecular cloud is highly sensitive to the physical properties
of the cloud. In order to obtain the chemical composition around a star forming region,
we carry out a two dimensional hydrodynamical simulation of the collapsing phase of a proto-star.
A total variation diminishing scheme (TVD) is used to solve the set of equations 
governing hydrodynamics. This hydrodynamic code is capable of mimicking evolution of the 
physical properties during the formation of a proto-star. We couple our reasonably large 
gas-grain chemical network to study the chemical evolution during the collapsing phase of a proto-star.
To have a realistic estimate of the abundances of bio-molecules in the 
interstellar medium, we include the recently calculated rate coefficients for 
the formation of several interstellar bio-molecules into our gas phase network. 
Chemical evolution is studied in detail by keeping grain at the constant
temperature throughout the simulation as well as by using the temperature variation
obtained from the hydrodynamical model. By considering a large gas-grain network
with the sophisticated hydrodynamic model more realistic abundances
are predicted. We find that the chemical composition
are highly sensitive to the dynamic behavior of the collapsing cloud, specifically on the
density and temperature distribution.
\end{abstract}


\end{frontmatter}

\section{Introduction}
The rate of discovery of molecules in the interstellar medium (ISM) has been increasing in
the past few years and today, over 170 confirmed molecules have been observed.
After the discovery of several molecules in the condensed phase,
it is now believed that the icy grains play a crucial role in enriching the ISM chemically.
Recent modeling results 
(Hasegawa, Herbst \& Leung 1992; Chakrabarti et al. 2006ab; Das et al. 2008b; Das \& Chakrabarti 2011)
also suggest that the gas-grain interaction needs to be appropriately modeled in order to mimic
the exact chemical evolution. However, the chemical evolution of any cloud is highly sensitive to the physical
properties of that cloud. It is therefore essential to study the physical properties of the cloud
at any instant for appropriate modeling. It is clear that the observed molecules must be synthesized
during the formation of the stars (i.e., in the proto-star phase).
Prior to the star formation, the interstellar chemistry
mainly followed by the gas-phase ion-molecule and neutral-neutral reactions leading to the formation of
small radicals and unsaturated molecules. During the collapsing phase, when the temperature
is cold enough and the density is much higher, most molecules accrete onto grains and form an icy
mantle (Vandishoeck et al., 1998). Recent theoretical work by Majumdar et al., (2013) showed that
how the computed spectrum differs in between the gas phase and ice phase. 
As the temperature starts to increase due to the formation of stars,
various species would return to the gas phase at the rate determined by their binding
energies with the grain surface. As a result, the gas phase chemical composition is modified by
the composition of the grain mantle.
In the present context, we carry out our investigation by revising our past
hydro-chemical model (Das et al. 2008a).
The hydrodynamical model is improved by introducing
two dimensional flow (instead of one dimensional model used by Das et al., 2008a) 
by the use of a scheme based on the
total variation diminishing (TVD) scheme (Harten 1983; Ryu et al., 1993). 

Our chemical model consists of gas phase chemical network as well as surface chemical network.
This large gas phase network includes the network of Woodall (2007).
In addition, we introduce 
some reactions which lead to the formation of bio-molecules by following
Chakrabarti \& Chakrabarti (2000ab), Majumdar et al. (2012, 2013).
We assume that the gas and the grains are coupled
through the accretion and thermal evaporation processes.
Most updated barrier energies are used for the grain chemistry 
network (Allen 1977, Das, Acharyya \& Chakrabarti 2010 and Das \& Chakrabarti 2011). Our
Chemical code is designed in such a way that any variation of the physical parameters such
as the density \& temperature are reflected in computing instantaneous rates.
The plan of the paper is the following. In the next section, the formulation of the
TVD scheme is discussed in details. Different aspects of our simulation results
are presented and discussed in Section 3, and finally in Section 4, we summarize our work.

\begin{figure}
\centering{
\vbox{
\psfig{figure=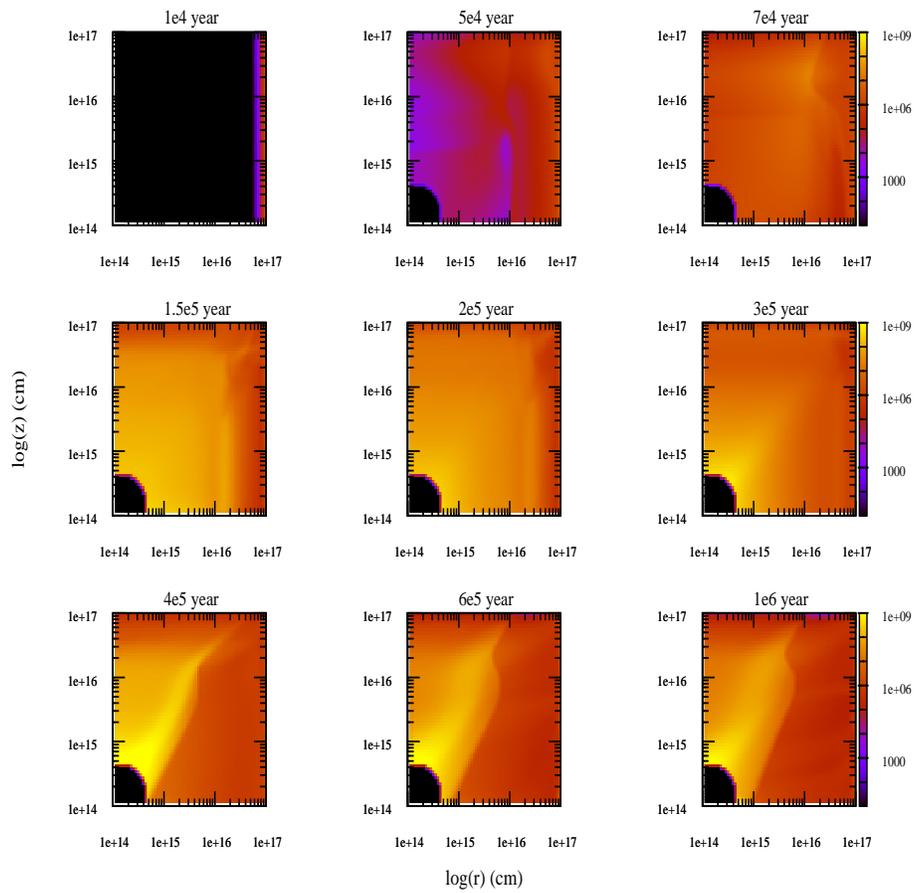,height=14cm,width=12cm,angle=-90}}}
\caption{\small{Density distribution of a collapsing cloud in the meridional plane. Time
is marked on each box. The color bar on the right can be used to get information about
number density.}}
\end{figure}

\section{Formulation of the TVD scheme}

In the past, we carried out the hydrodynamic simulation for a spherically symmetric
gas cloud (Das et al. 2008a). 
The code was developed to study the dynamic behaviour of
a spherically symmetric isothermally (T=10K) collapsing cloud. However, in order to study 
the chemical evolution more realistically, it is necessary to study the dynamic behavior 
in a two dimensional flow
which includes rotation. In the present paper, we implement these aspects.

We solve the following hyperbolic system of conservation equations for
a collapsing interstellar cloud:

\begin{equation}
\frac{\partial \rho}{\partial t}+\frac{\partial}{\partial x_k}(\rho u_k)=0,
\end{equation}

\begin{equation}
\frac{\partial (\rho u_i)}{\partial t}+\frac{\partial}{\partial x_k}
(\rho u_i u_k+p \delta_{ik})=0
\end{equation}

\begin{equation}
\frac{\partial E}{\partial t}+ \frac{\partial}{\partial x_k}[(E+p)u_k]=0,
\end{equation}
where, $E=\frac{p}{(\gamma-1)} + \frac{\rho {u_k}^2}{2}$ is the total energy per unit volume and the rest
of the variables have their usual meanings. We use the TVD scheme to solve the above hydrodynamic
equations in cylindrical coordinates (r, $\Phi$, z). Here we denote $r$ coordinate
by $x$ and $\Phi$ by $y$. Harten's TVD scheme (Harten 1983)
is an explicit, second order Eulerian finite difference scheme, which solves a hyperbolic system of the
conservation equations. The key merit of this scheme is to achieve a high resolution.
This scheme is relatively simple to program compared to the other high
accuracy numerical schemes and require less CPU time. Here we assume there are no
variations along the $y$ ($\Phi$) direction, so the code is actually two dimensional in nature.

For the inclusion of entropy, we introduce another conservative quantity,
$S=\frac{P}{\rho^{\gamma-1}}$. Combining the energy equation (Eqn. 3)
with the mass conservation equation (Eqn. 1), modified entropy equation becomes
\begin{equation}
\frac{\partial S}{\partial t}+\frac{\partial}{\partial x_k}(S u_k)=0.
\end{equation}
The mass and momentum conservation equations (Eqn. 1 and Eqn. 2 respectively),
along with the above modified entropy equation can be written in the vector form as the following:
\begin{equation}
\partial_t q + \partial_x F_x + \partial_y F_y + \partial_z F_z = M,
\end{equation}
where, $M$ is the source vector, $F_x(q)$, $F_y(q)$, $F_z(q)$ are the flux functions with vector q.
The Jacobian matrices, $A_x(q)=\frac{\partial F_x}{\partial q},
A_y(q)=\frac{\partial F_y}{\partial q}$, and $A_z(q)=\frac{\partial F_z}{\partial q}$,
are formed by using the flux functions. All the vectors are denoted as follows:
\begin{figure}
\vskip 1cm
\centering{
\vbox{
\psfig{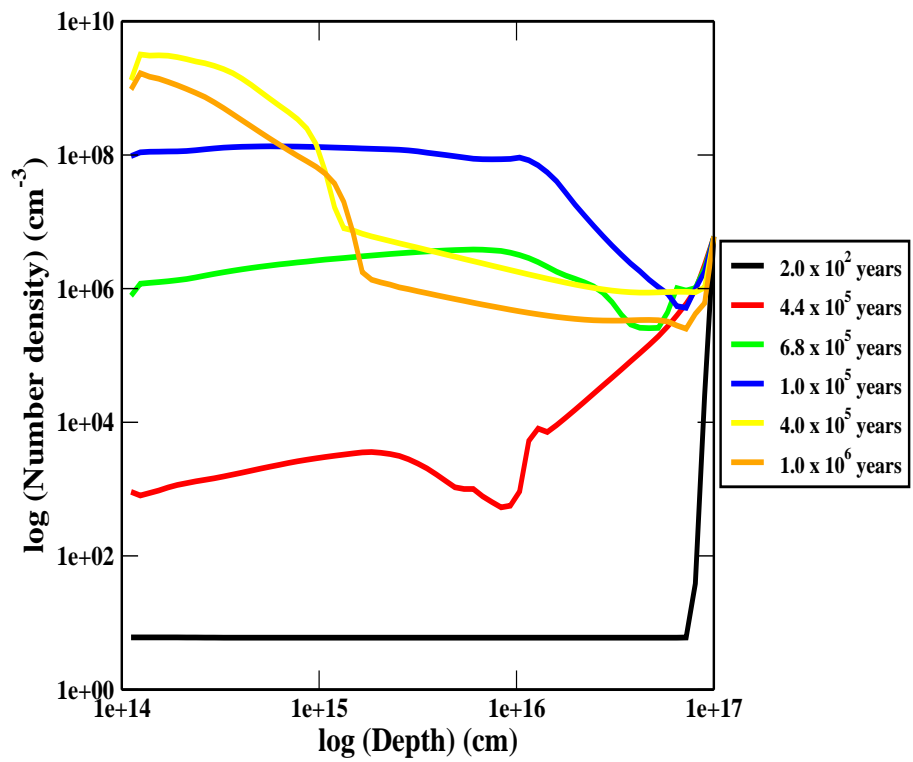}}}
\caption{\small{Time evolution of the number densities for the plane at z= 4.53$\times 10^{14}$ cm.}}
\end{figure}
%\begin{equation}
$$
q = \left(
   \begin{matrix}
    \rho \cr
    \rho v_x \cr
    \rho v_y \cr
    \rho v_z \cr
     S
    \end{matrix}
    \right),\qquad
F_x = \left(
      \begin{matrix}
      \rho v_x \cr
      \rho{v_x}^2 + S \rho^{\gamma -1} \cr
      \rho v_x v_y \cr
      \rho v_x v_z \cr
      S v_x
     \end{matrix}
     \right),\qquad
F_y = \left(
      \begin{matrix}
      \rho v_y \cr
      \rho v_x v_y \cr
      \rho{v_y}^2 + S \rho^{\gamma -1} \cr
      \rho v_y v_z \cr
      S v_y
      \end{matrix}
      \right),\qquad
$$
%\end{equation}
\begin{equation}
F_z = \left(
      \begin{matrix}
      \rho v_z \cr
      \rho v_x v_z \cr
      \rho v_y v_z  \cr
      \rho{v_z}^2 + S \rho^{\gamma -1} \cr
      S v_z
      \end{matrix}\\
      \right),\qquad
M=\left(
\begin{matrix}
0 \cr
          \rho {v_y}^2/x \cr
          \rho v_x v_y/x \cr
          0\cr
          0\cr
\end{matrix}
\right)
\end{equation}
%and $M$ is the source vector.
We calculate all the eigen values using the characteristic equation or secular equation and the values are,
$$
%a_1 = \frac {-\left[-c^2 \gamma \rho^4 + c^2 \gamma^2 \rho^4 + \gamma^2 \rho^{\gamma + 2} S \right ]^\frac{1}{2} + \gamma \rho^2 v_x}{\gamma \rho^2}         = v_x-c,
a_1 = v_x-c,
$$
$$
a_2 = v_x,
$$
\begin{equation}
a_3 = v_x,
\end{equation}
$$
a_4 = v_x,
$$
$$
%a_5 = \frac {\left[-c^2 \gamma \rho^4 + c^2 \gamma^2 \rho^4 + \gamma^2 \rho^{\gamma + 2} S \right ]^\frac{1}{2} + \gamma \rho^2 v_x}{\gamma \rho^2}         = v_x+c.
a_5 = v_x+c.
$$
\begin{figure}
\vskip 1cm
\centering{
\vbox{
\psfig{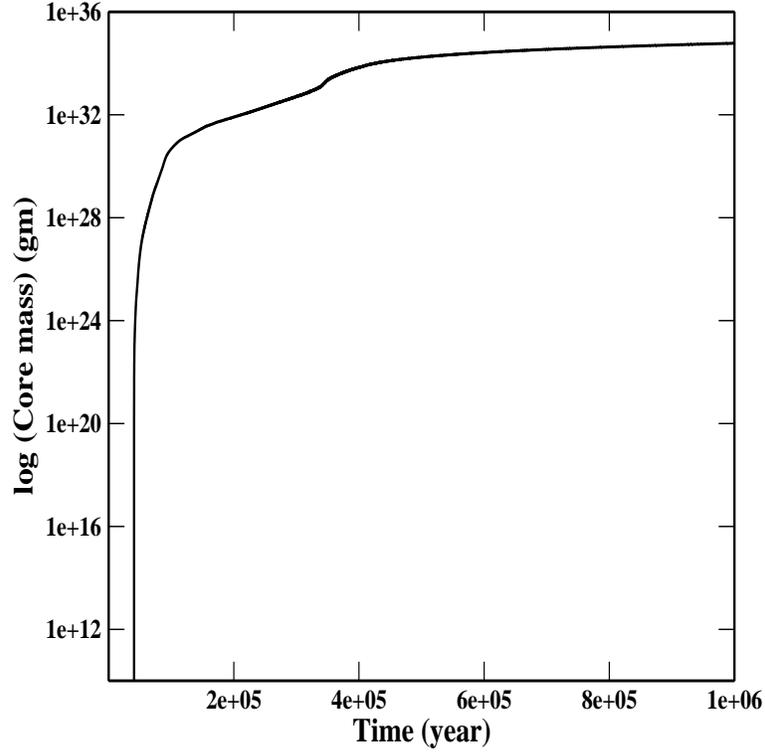}}}
\caption{\small{Time evolution of the core mass is shown.}}
\end{figure}
%\subsection{Determination of Eigenvectors}
%
%\subsubsection{Right Eigenvectors}
To find the eigenvectors of $A_x$ with respect to the eigenvalue a$_1$, we have considered
the following eigenvalue equation,
\begin{equation}
(A_x-a_1 I) R_1 = 0,
\end{equation}
where, $I$ is the Unit matrix.
%\begin{equation}
% \left(
%   \begin{matrix}
%    c-v_x & 1 & 0 & 0 & 0  \cr
%     \frac{c^2 (\gamma - 1)}{\gamma} - v_x^2 & v_x + c & 0 & 0 & \rho^{\gamma - 1} \cr
%      -v_x v_y & v_y & c & 0 & 0 \cr
%      -v_x v_z & v_z & 0 & c & 0 \cr
%       \frac{- S v_x}{\rho} &\frac{S}{\rho}  & 0 & 0 &  c
%    \end{matrix}
%    \right)\qquad
%   \left(
%     \begin{matrix}
%      x_1 \cr
%      x_2 \cr
%      x_3 \cr
%      x_4 \cr
%      x_5
%     \end{matrix}
%     \right)\qquad
%  =  \left(
%     \begin{matrix}
%      0 \cr
%      0 \cr
%      0 \cr
%      0 \cr
%      0
%     \end{matrix}
%     \right).\qquad
%\end{equation}
By solving the above, the right eigenvector corresponding to the eigenvalues are calculated.
%\begin{equation}
%R_1 =  \left(
%     \begin{matrix}
%      x_1 \cr
%      x_2 \cr
%     x_3 \cr
%      x_4 \cr
%      x_5
%     \end{matrix}
%     \right)\qquad
%    = \left(
%     \begin{matrix}
%      1 \cr
%      v_x-c \cr
%      v_y \cr
%      v_z \cr
%      \frac{S}{\rho}
%     \end{matrix}
%     \right).\qquad
%\end{equation}
%Other right eigenvectors corresponding to the eigenvalues $v_x$ , $v_x$ , $v_x$, $v_x$ + c are obtained respectively as following,
%$$
%R_2 =  \left(
%     \begin{matrix}
%      0 \cr
%      0 \cr
%      1 \cr
%      0 \cr
%      0
%     \end{matrix}
%     \right),\qquad
%$$
%\begin{equation}
%R_3 =  \left(
%     \begin{matrix}
%      1 \cr
%      v_x \cr
%      v_y \cr
%      v_z \cr
%      -\frac{(\gamma - 1)S}{\rho}
%     \end{matrix}
%     \right),\qquad
%\end{equation}
%$$
%R_4 =  \left(
%     \begin{matrix}
%      0 \cr
%      0 \cr
%      0 \cr
%      1 \cr
%      0
%     \end{matrix}
%     \right),\qquad
%$$
%$$
%R_5 =  \left(
%     \begin{matrix}
%      1 \cr
%      v_x + c \cr
%      v_y \cr
%      v_z \cr
%      \frac{S}{\rho}
%     \end{matrix}
%     \right).\qquad
%$$
\begin{figure}
\vskip 1cm
\centering{
\vbox{
\psfig{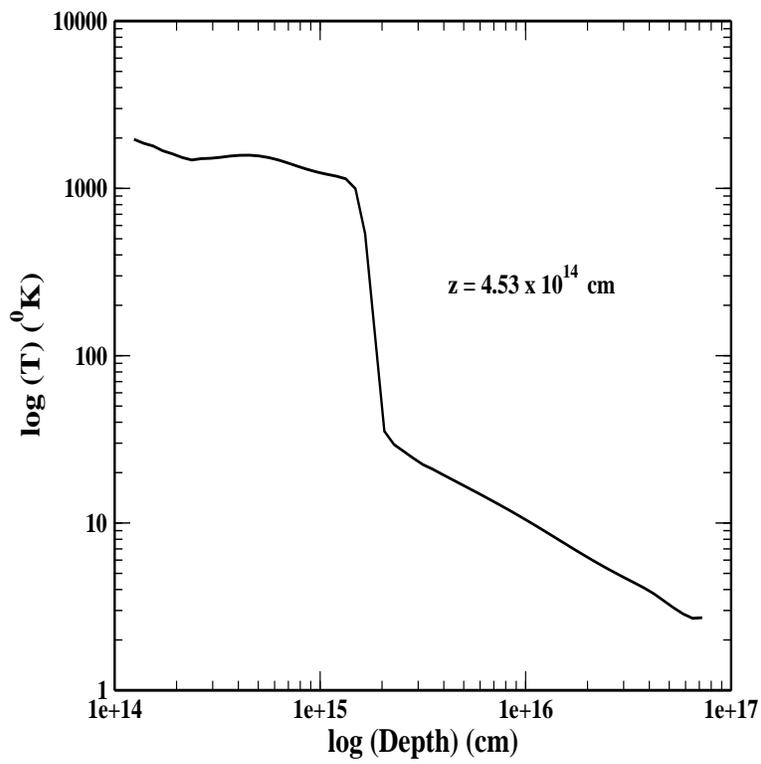}}}
\caption{\small{Temperature distribution at z=4.53 $\times 10^{14}$ cm plane}}
\end{figure}
%\subsubsection{Left Eigenvectors:}
The set of left eigenvectors ([L]) are determined from the inverse of the right eigenvectors ([R]):
\begin{equation}
[L] = [R]^{-1}.
\end{equation}
%Thus the left eigenvectors, which are orthonormal to the right eigenvectors are given by,
%$$
%L_1 = \left[\frac{(\gamma - 1)}{2 \gamma} + \frac{v_x}{2c}  ,- \frac{1}{2c^2},0,0,\frac{\rho}{2 \gamma S}\right],
%$$
%$$
%L_2 = [- v_y, 0,1,0,0],
%$$
%\begin{equation}
%L_3 = \left[ \frac{1}{\gamma},0,0,0,- \frac{\rho}{\gamma S}\right],
%\end{equation}
%$$
%L_4 = [- v_z, 0,0,1,0],
%$$
%$$
%L_5 = \left[\frac{(\gamma - 1)}{2 \gamma} - \frac{v_x}{2c}  , \frac{1}{2c^2},0,0,\frac{\rho}{2 \gamma S}\right].
%$$
We use the Roe approximate Riemann solution (Roe, 1981) to get the averaged values of the
physical quantities at the grid boundaries:
$$
\rho_{i+\frac{1}{2}}=\frac{\rho_i+\rho_{i+1}}{2},
$$

$$
v_{x,i+1/2}=\frac{\sqrt{\rho_i}v_{x,i}+\sqrt{\rho_{i+1}}v_{x,i+1}}
{\sqrt{\rho_i}+\sqrt{\rho_{i+1}}},
$$

$$
v_{y,i+1/2}=\frac{\sqrt{\rho_i}v_{y,i}+\sqrt{\rho_{i+1}}v_{y,i+1}}
{\sqrt{\rho_i}+\sqrt{\rho_{i+1}}},
$$

\begin{equation}
v_{z,i+1/2}=\frac{\sqrt{\rho_i}v_{z,i}+\sqrt{\rho_{i+1}}v_{z,i+1}}
{\sqrt{\rho_i}+\sqrt{\rho_{i+1}}},
\end{equation}

$$
S_{i+1/2}=\frac{\sqrt{\rho_i}{S_i}+\sqrt{\rho_{i+1}}S_{i+1}}
{\sqrt{\rho_i}+\sqrt{\rho_{i+1}}},
$$

$$
c_{i+1/2}=\gamma S_{i+1/2} {\rho^{\gamma-2}_{i+1/2}}.
$$
To incorporate the self-gravity in this code, we introduce gravitational potential.
The source function including the self-gravity term becomes,
\begin{equation}
M=\left(
\begin{matrix}
0 \cr
          \rho {v_y}^2/x - T_x\cr
          \rho v_x v_y/x \cr
          - T_z\cr
          - (T_x x+T_z z)\cr
\end{matrix}
\right),
\end{equation}
where, $\phi$ is the potential energy per unit mass,
$T_x=\rho \frac{d \phi}{dx}$ and $T_z=\rho \frac{d \phi}{dz}$ are
the components along the x and z directions respectively.
$\frac{d\phi}{dx}$ and $\frac{d\phi}{dz}$ are the gravitational force components
per unit mass along the x and z directions
respectively. Gravitational potential energy per unit mass ($\phi$), is calculated as follows;
$$
\phi=\phi_s+\phi_c ,
$$
where, $\phi_s$ is the potential energy per unit mass of the cloud and
$\phi_c$ is the potential energy per unit mass due to the central core mass (M$_c$(t)) at any time t.
M$_c$(t) is increasing with time as follows:
\begin{equation}
M_c(t)=M_c(t)+M_s(t),
\end{equation}
where, $M_s(t)$ is the amount of mass dumping by the cloud at the instant $t$. This amount depends on the
physical properties of the grid from where they are contributing to the core. In our code,
we assume that any mass going inside the threshold radius ($R_c$) at any instant $t$ will be treated as $M_s(t)$.
\begin{figure}
\vskip 2cm
\centering{
\vbox{
\psfig{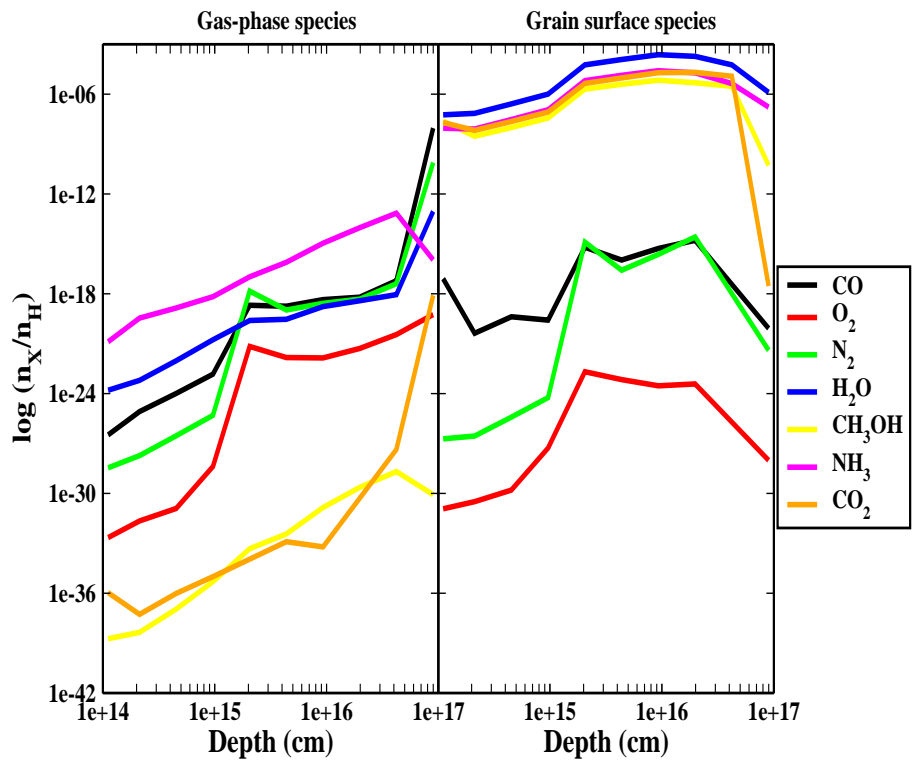}}}
\caption{\small{Abundances of various major species in the gas and grain phases as the depth 
(x-coordinate) is varied at z=4.53 $\times 10^{14}$ cm plane. 
Note that most of the region of cloud (gas phase/grain phase), 
$H_2O$ dominates.}}
\end{figure}

\begin{figure}
\vskip 1cm
\centering{
\vbox{
\psfig{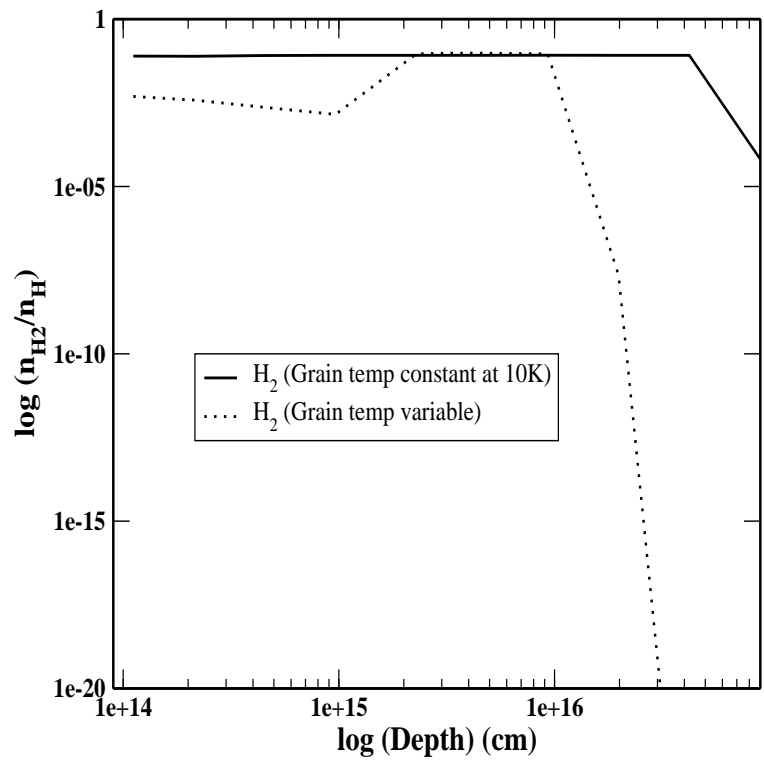}}}
\caption{\small{Depth dependency of gas phase H$_2$ abundance is shown for
different depth at z=4.53 $\times 10^{14}$ cm plane.
Solid curve is for the case where grain temperature is assumed to be constant at 10K 
and dotted curve is for the case where grain temperature varies.}}
\end{figure}

\subsection{Effects of Heating and Cooling}
There are various physical processes which could affect the physical properties of the ISM, such as, 
heating and cooling of the interstellar medium. 
Molecular clouds are generally cold. Tilens (2005) summarizes calculated cooling rates for 
interstellar molecular clouds. According to Tilens (2005), Cooling rates basically depends upon the 
abundances or, more precisely, on the ratio of the abundance to the velocity gradient,
which sets the optical depth of the transitions involved. 
At low densities, CO dominates the cooling because of its high
abundance in molecular clouds. Around a density of 10$^3$ cm$^{-3}$ contribute to the cooling. 
Following Tilens (2005), we have included the
cooling rate as a function of the $H_2$ number density of the cloud.

According to (Tielens 2005), the low energy cosmic rays (CR) ($\sim$ 1 - 10 MeV) are the most 
efficient for ionizing and heating the gas. 
Here also, we consider only the effects of the cosmic rays for the process of gas heating. According to (Tielens 2005) 
, Cosmic ray heating rate can be represented as the following:
\begin{equation}
R_{CR}=n \zeta_{CR}E_h(E,xe),
\end{equation}
where, $\zeta_{CR}$ is the total cosmic ray ionization rate ($3 \times 10^{-16}\ s^{-1}$ by assuming
primary ionization rate $2 \times 10^{-16} s^{-1}$), $n$ is the number
density in the units of cm$^{-3}$ and $E_h(E,x_e)$ is the average heat deposited per primary
ionization. For the low degree of ionization, $E_h(E,x_e)\sim 7 \ eV$. In this case, 
the above equation becomes,
\begin{equation}
R_{CR}= 3 \times 10^{-27} n \frac{\zeta_{CR}}{2 \times 10^{-16}} \ erg \ cm^{-3} \ s^{-1}.
\end{equation}
During various chemical reactions, heats are released/absorbed. But we have noticed that these rates are
comparatively much lower than the other heating and cooling terms discussed above. So in our simulation,
we neglect effects of energy released or absorbed due to chemical reactions.

\section{Results and Discussion}
\begin{table}
\centering
\caption{Units used in the TVD code.}
\begin{tabular}{ll}
\hline\hline
Parameter&Value\\
\hline\hline
Length&$10^{17}$cm\\
Velocity&$3 \times 10^{10}$cm/sec\\
Density&$10^{-17}$gm/cm$^{-3}$\\
Time&$3.33 \times10^6$second\\
\hline\hline
\end{tabular}
\end{table}

\begin{table}
\centering
\caption{Initial and boundary values in c.g.s. units.}
\begin{tabular}{ll}
\hline\hline
Parameter&Value\\
\hline\hline
Outer radius&$10^{17}$cm\\
Inner radius&$10^{14}$cm\\
Initial Density&$10^{-23}$gm/cm$^3$\\
Courant factor&0.1\\
Initial Core mass& 0 gm\\
Velocity at outer x boundary &-2.5 $\times 10^3$ cm/sec\\
Angular rotation at outer x boundary &$10^{-14}$ sec$^{-1}$\\
Density at the outer x boundary&$10^{-17}$gm/cm$^{3}$\\
$\gamma$&5/3\\
Number of grids&$64 \times 64$\\
Threshold radius&$4.06 \times 10^{14} $cm\\
\hline \hline
\end{tabular}
\end{table}

We now present the results of the simulations for a concrete case.
In Table 1, the parameters which are used to non-dimensionalize the TVD code are tabulated and in Table 2, the parameters which are used here are written in c.g.s. units.
We consider an interstellar cloud having a size of $10^{17}$ cm
(i.e., $0.03$ parsec) and divide the entire cloud into the 64$\times$64 logarithmically equal
spaced grids along the x and z directions respectively.
This may appear to be a low resolution run. However, we find little difference in
the average thermodynamic properties of cloud between a $16 \times 16$ run and a $64 \times
64$ run. Since the main emphasis of this paper is to study the chemical evolution of a
number of species which depends on the average properties of the cloud,
any further refinement of the mesh may not be necessary at this stage.

The cloud is assumed to be axisymmetric.
Initially (at time $t = 0$), we assume that the cloud contains some mass
(having density $10^{-23}$ gm$/$cm$^{-3}$
at each grid location).
From the outer boundary, we start to inject matter at a constant rate with a density $10^{-17}$ gm cm$^{-3}$
, inward radial velocity  $-2.5 \times 10^3$ cm/sec (-ve sign indicates the inward flow) and angular 
rotation 10$^{-14}$ sec.
In the TVD code, we use non-dimensional quantities with the density at the outer boundary
as the unit of density and the length of the computational zone along the x-axis as the unit of length.
A sink is kept at the center of the cloud and it is assumed that any
matter going inside a particular radius (the threshold radius, $R_c$) contributes to the core mass.
Time step to advance the global time is calculated by using Courant-Friedrichs-Lewy
(hereafter, Courant) condition. To be on the safer side even smaller time steps are used (Courant factor $0.1$).

\begin{figure}
\vskip 1cm
\vskip 3cm
\centering{
\vbox{
\psfig{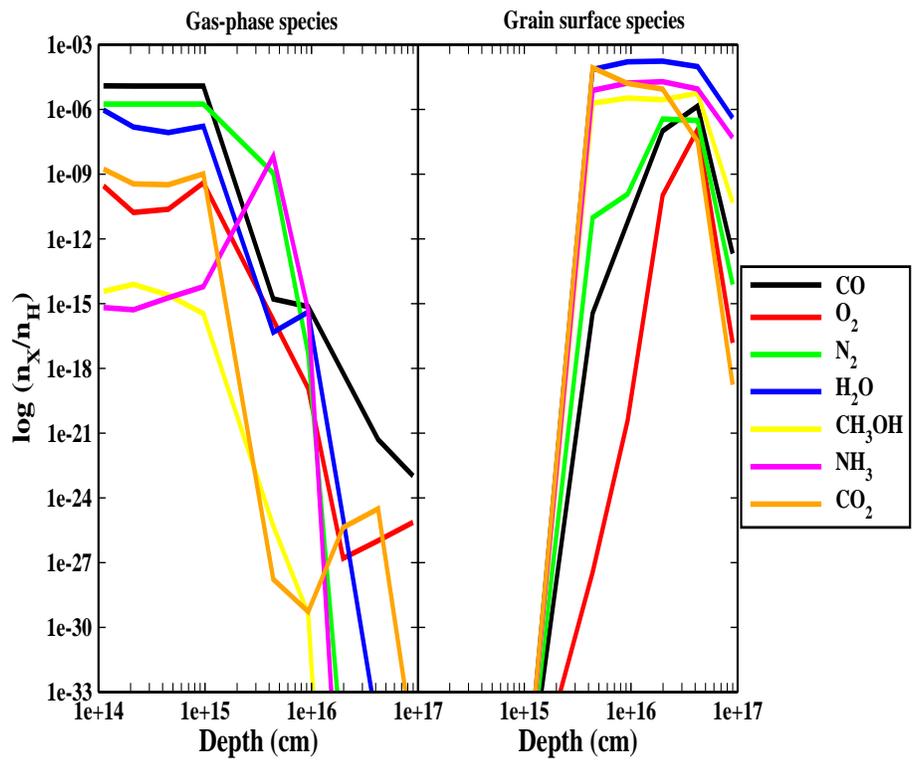}}}
\caption{\small{Depth dependence of various simple species when the variation of
grain temperature is also included.}}
\end{figure}
Figure 1 represents the time evolution of the density distribution of a 
collapsing cloud in the meridional plane.
The density of the cloud increases with time as the matter is accreted.
On the right side of the panel, a colour box is shown to indicate the colour codes of the number densities
in the logarithmic scale. For instance,
black corresponds to a very low density and yellow corresponds to a very high density (marked on the right).
We note that with time, the grid is filled up with a high density matter.
Meanwhile, with time, the matter content of the innermost grid at the
center (indicated by a large black box) also increases.

The black sphere at the centre does not mean the core. Our actual core is point-like
at the center. However, we have assumed a sink radius (the black box) which is sucking the
matter from the cloud and transferring it into the core. This is required for the purpose of
our numerical scheme to realize the effect of the central object. It should be kept in mind
that we were dealing with a logarithmic scale here. So the apparently large looking sink is really very
mini scale in size (size of the sink is 4.06 $\times 10^{14}$ cm and the whole size of the cloud is $10^{17}$ cm).

To understand Fig. 1 more clearly, in Fig. 2, we show the time evolution
of number densities at a plane which is situated at a height of z=$4.53 \times 10^{14}$cm.
The reason behind choosing these particular height is
that the innermost grid location at this height is just beyond R$_c$ and thus we see
the density distribution around the innermost grid of this plane as well.
Along the x axis, we plot the distance along the x direction and along the z axis, we plot the density at each grid.
The right panel indicates time at which the density distributions are plotted.
To have an idea about how the mass of the core increases with time, in
Fig. 3, the time evolution of the total amount of accreted matter through the
threshold radius is shown. As the density around $R_c$ increases, the accretion rate
from the $R_c$ also increases and as a result, the core mass increases rapidly.
In the present work, we assume that the gas is adiabatic. So the temperature is also changing dynamically.
In Fig. 4, the temperature distribution throughout a plane (at a height of z=4.53$\times 10^{14}$cm) of a cloud
is shown at the end of the simulation (i.e., after 1$\times 10^6$ year).
From Fig. 4, it is clear that the temperature increases as we enter deeper inside the cloud.

In order to perform a self-consistent study, we assume that the gas and the grains are coupled
through the accretion process and the
thermal evaporation processes. We assume that the species are physisorbed onto the dust
grain (classical size grain $\sim 1000 A^{\circ}$) having the grain number density
$1.33 \times 10^{-12} \times n$, where $n$ is the concentration of H nuclei in all forms.
Following, Hasegawa et al. (1992),we assume that there are 156 surface reactions
among the 118 neutral surface species in our grain chemistry network.
Following Das \& Chakrabarti (2011) and references therein,
all the updated interaction energies are used in our grain surface network.
In addition to a large grain chemistry network, a large network of gas phase chemistry 
following Woodall et al., (2007)
is implemented. Following Majumdar et al., (2012), we add up some more gas phase reactions 
in our network to have an educated
estimation about the biologically important interstellar species. Majumdar et al., (2012)
performed a quantum chemical calculation
to find out the reaction rates for the formation of some bio-molecules.

In our hydrodynamical model, we have 64$\times$64 numerical grids. So in order to study the
chemical evolution of the entire cloud, we need to study the chemical evolution at each numerical grids.
\begin{figure}
\vskip 1cm
\vskip 3cm
\centering{
\vbox{
\psfig{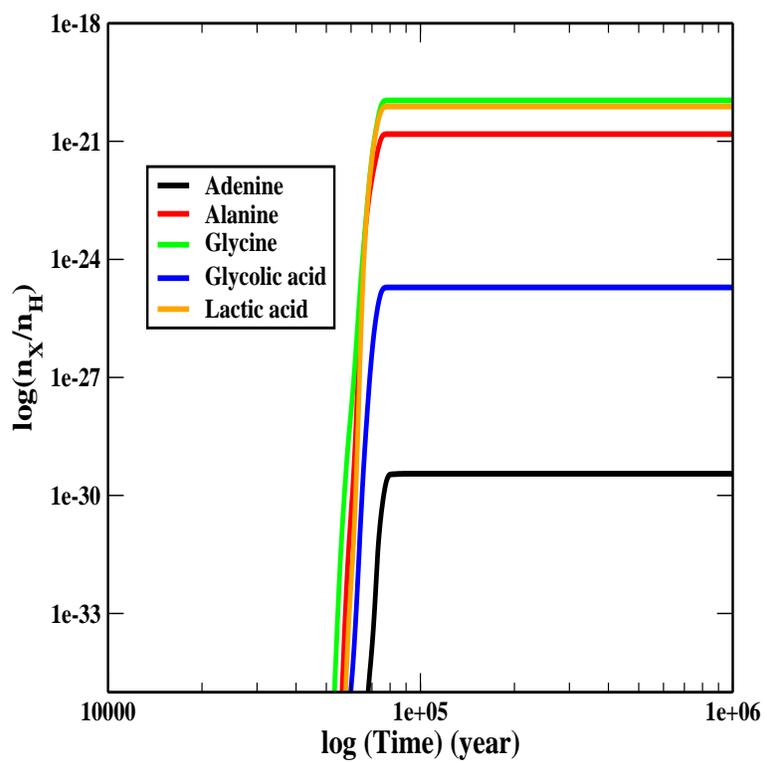}}}
\caption{\small{Time evolution of some of the pre-biotic molecules.}}
\end{figure}
In Fig. 5, we show the variation of the final abundances of some of the major interstellar
gas-grain species for the plane at the height of 4.53 $\times 10^{14}$ cm).
Here, we assume that the grain temperature remains constant at 10K and vary
gas temperature according to the outcome of the hydrodynamic code.
Around the deep inside the cloud, density is much higher but it is also
evident that the temperature is also much higher (Fig. 4). Abundances of several gas phase 
species are decreasing due to the heavy depletion to the grain surface. Since for the sake of
simplicity, here we have assumed that grain temperature remain invariant at 
10K during the life time of the molecular cloud, most of the species locked into the
grain surface due to the tight binding with the grain surface.
Gas phase abundance around the deep inside the cloud also decreases
due to the several favourable fragmentation reactions at the high temperatures.
Stable species like H$_2$O, CH$_3$OH, CO$_2$, which are the major constituents
of the grain mantle producing efficiently around the intermediate region of the 
collapsing cloud. Throughout the cloud, water is the most dominating molecule in the
gas/grain phase.

\subsection{Temperature dependent study}
So far, we have carried out the simulations by keeping the grain temperature to be constant at T$=10$K.
Gas phase temperature were taken from the hydrodynamic code. We have tested our case by assuming
the actual temperature variations for the grains also.
Consideration of the temperature effect, heavily affects the final results.

The molecular hydrogen is very abundant in the ISM. But this huge abundance cannot be
explained unless we invoke the grain chemistry (Chakrabarti et al. (2006ab)).
Gas phase H atoms land on the dusty grains and
produce molecular hydrogen via surface migration or Eley-Rideal method (direct accretion of one H on the
top of another H atom) and finally they are evaporated from the grain surface and contribute to 
the gas phase abundances.
For the better understanding, in Figure 6, we have shown the depth dependence of the 
molecular hydrogen abundance at a height of z=4.53$\times10^{14}$cm.
It is well known that around the 10-20K, H$_2$ formation efficiency is maximum (Biham et al., 2001,
Chakrabarti et al., 2006ab).
It is clear from Fig.6 that in case of the solid curve, H$_2$ is forming very efficiently around 
deep inside the cloud, whereas for the
dotted curve it is producing around the intermediate region where temperature is much lower (10-20K).
This figure basically explains how the conversion of atomic hydrogen to the hydrogen molecule
occurs at different depth of our simulated cloud.
For the solid curve, we assume that grain temperature remain constant at 10K and
for the dotted curve, we consider more realistic case where the temperature variation of 
grains are also taken into account. 
In reality dotted curve is much more convenient to use since deep inside the cloud
though the density is favourable for the formation of molecular hydrogen,
temperature is much higher, as a result H$2$ production
is not significant due to the short life time of the H atom on the grain surfaces.

In Figure 7, we show the depth dependence of some of the major gas-grain species by considering
the temperature dependency in the grain phase.
For the better understanding of the difference between the two considerations (temperature dependency
of grain and grain kept at 10K), here also, we consider the same plane 
at z=4.53$\times 10^{14}$cm as the earlier.
Comparing Fig. 7 with the Fig. 5, we can clearly see that deep inside the cloud, the production over
the interstellar grain is insignificant in Fig. 7. This is because, deep inside the cloud, the temperature
is higher and all the molecules are likely to be evaporated from the grain surface and indeed
the grain mantles themselves may evaporate, at least partly. Hence, around that region, the chemical
enrichment of the interstellar medium is continued by the gas-phase chemistry alone. We have noticed
that when we are considering the temperature variation, the effect of grain is found to be negligible
inside of $\sim 2\times 10^{15}$ cm. However, outside this radius, the temperature is cool
enough for the production of several interstellar molecules on the grain surface. At the
same time, as we go out from the central region, the density decreases, and the production
on the grains significantly decreases.
In case of the gas phase chemistry, deep inside the cloud, the temperature
is higher but as the density is also higher,  abundances are enhanced. In reality,
the situation is very complex. Due to higher temperatures, the molecules dissociate
easily into much smaller components. However, since the density is much
higher also, they might be able to re-united again. So there is always a competition going on
between the two effects: one due to density and and other due to temperature.

For example, in case of Fig. 5, gas phase CO molecules are rapidly decreasing due to the heavy
depletion on the grain surface, whereas in case of Fig. 7, due to the high temperature of
the grain surface, CO molecules are unable to reside on the grain surface. As a result, their
abundance in the gas phase increases. Production of surface species are noticed to be 
favourable around the low temperature region (8-30K). 
From the temperature distribution (Fig. 4) of the cloud, it is evident that intermediate 
region of the cloud ($\sim 2\times 10^{15}-1.5 \times 10^{16}$cm,) 
is also in the low temperature regime (8-30K), so the production of surface
related species are more favourable around the intermediate region of the cloud. 
From Fig. 7, it is also clear 
that production of the grain surface species decreases as we are going further out, especially
because density is reduced with distance. At the much lower temperatures ($<$8K), surface
species also lost its mobility, as a results production also hindered in the grain phase also.

We believe that inside a molecular cloud bio-molecules might also be formed due to very complex 
and rich chemical process. Production of amino acids, nucleobases, carbohydrates and other basic 
compounds can possibly start from these bio-molecules. For the first time 
Chakrabarti \& Chakrabarti (2000ab) made an attempt to study the formation of 
bio-molecules (adenine, alanine, glycine, glycolic acid and lactic acid) during the 
collapsing phase of a proto-star.
They realized that major obstacle for studying the evolution of these interstellar 
bio-molecules are the lack of 
adequate knowledge of the rate co-efficients of various reactions which are taking place.
In order to obtain more realistic abundances of interstellar bio-molecules, Majumdar et al., (2012)
carried out quantum chemical simulation to find out the reaction rate coefficients for the
formation of these interstellar bio-molecules.
Here, we have included the chemical network for the formation of these bio-molecules 
by following Chakrabarti et al., (2000ab) \& Majumdar et al., (2012).
 
In Fig. 8, we have shown the chemical evolution of some important bio-molecules 
at a grid location $X=9.3 \times 10^{15}$cm, $Z= 4.53 \times 10^{14}$cm,
where temperature is around 10K, which is favourable for the production of complex molecules.
Computed abundances of these molecules are 
$1.1 \times 10^{-20}, 1.53 \times 10^{-21}, 3.57 \times 10^{-30}, 7.66 \times 10^{-21}, 
1.94 \times 10^{-25}$ for glycine, alanine, adenine, lactic acid \& glycolic acid respectively.
Majumdar et al., (2013) made a rigorous attempt to identify the precursor of adenine, alanine \& glycine 
which could be observed in the ISM. 
According to them NH$_2$CN and HCCN are the precursors of adenine, C$_2$H$_3$ON is the precursor 
of glycine and C$_3$H$_5$ON is the precursor of alanine. They also reported 
their respective infrared and electronic absorption spectra.
Here, our computed abundances are well below the observational limit, thus as Majumdar et al., (2013), 
we are also proposing to predict the abundances of these molecules by observing its pre-cursor molecules.
Spectral information of these pre-cursor molecules are already discussed in detail in 
Majumdar et al., (2013). To set a observational guidelines for predicting the
abundance of these bio-molecules, as like Das et al., (2013), 
we are now trying to report the rotational spectral information of these pre-cursor molecules 
elsewhere soon in the format of JPL/CDMS catalog.

\subsection{Comparison with previous models}
Main emphasis of the present paper is upon the chemical evolution of a more realistic cloud
by considering a two dimensional hydrodynamical flow which includes rotation and heating \& cooling
of the ISM whereas our earlier paper (Das et al., 2008a) 
mainly focuses on the chemical evolution of a spherically symmetric isothermal collapsing cloud.
Our present chemical model is more upto dated
as here we have used UMIST 2006 database (Woodall et al., 2007) for our gas phase chemical network 
and have included recently computed rate coefficients for the formation of several bio-molecules
(Majumdar et al., 2012).  In our earlier model, 
we considered grain chemistry only for the formation of molecular hydrogen, here we are
using a large surface chemical network (Hasegawa, Herbst \& Leung, 1992, Das, Acharyya \& Chakrabarti 2010
and Das \& Chakrabarti 2011) to self-consistently study the chemical evolution
of a collapsing cloud. In our earlier paper as we have considered an isothermal cloud, chemical
evolution was mainly dependent upon the density of any region, whereas, in the present model,
temperature is also an important parameter for deciding the degree of chemical enrichment.
To avoid complexity in this paper, we have studied the chemical evolution at a particular height
(Z$=4.53 \times 10^{14}$cm), but in principal we could have studied it for the whole cloud. This
is out of scope of this paper and we are now in preparation to report it elsewhere.
Since the chemical evolution heavily dependent upon the physical parameters (density \& temperature),
abundances of all the species would expected to be different along different region. 

Hasegawa, Herbst \& Leung (1992) prepared a gas-grain model and studied the
time evolution of the chemical species for a steady state cloud ($n=2 \times 10^4$ cm$^{-3}$, T=10K
and A$_V$=500). 
Stantcheva \& Herbst (2004) prepared models of gas-grain chemistry in interstellar cloud cores
($n=2 \times 10^4$ cm$^{-3}$, T=10-20K) with a stochastic approach to surface chemistry. 
For both the
cases it was observed that as time is evolving, CO molecules are heavily depleted from 
the gas phase and the abundance of CO and its related surface species (CO$_2$, H$_2$CO, CH$_3$OH) 
gradually increases over the time being. Our model mainly differs due to the physical properties
of the cloud considered here. In this paper, we are adopting the physical properties from the outcome of our
updated hydrodynamical model, whereas they considered a static cloud condition. 
Here, Fig.7, shows that deep inside the cloud,
abundances of gas phase CO molecules are much higher whereas in Das et al., (2008a) it was shown that
deep inside the cloud gas phase CO molecules are depleted to the grain surface. 
In Das et al., (2008a), it was assumed that cloud remain in isothermal stage during the
collapsing phase. So, as we are going deep inside the cloud, since the density is increasing,
probability of freezing to the grain surface increases simultaneously.
But in reality, as we are going inside a collapsing cloud, its temperature increases gradually (Fig. 4),
and the probability of freezing gradually diminishes. Since in Fig. 7, effect of temperatures are 
included, it is showing apparently the real depth dependency of a collapsing cloud.

In Das et al., (2008a), abundances of alanine \& glycine were computed 
by assuming an average reaction rate coefficients $\sim 10^{-10}$ cm$^3$s$^{-1}$. Computed 
abundances of alanine and glycine from Das et al., (2008a) was  $2.3-8.3 \times 10^{-17}$ 
and $1.7-2.9 \times 10^{-14}$ respectively. With the modified rate coefficients, Majumdar et al.,
(2013) predicted the peak abundance of alanine and glycine to be $8.9 \times 10^{-18}$ and 
$1.96 \times 10^{-17}$ respectively. Our computed abundances of the alanine and glycine are 
$1.53 \times 10^{-21}$ and $1.1 \times 10^{-20}$ respectively. 
Chakrabarti \& Chakrabarti (2000a) predicted the abundances of adenine to be $6.35 \times 10^{-11}$
whereas Majumdar et al., (2012, 2013) predicted adenine abundance to be $4.4 \times 10^{-25}$. In our
present context adenine abundance found to be $3.57 \times 10^{-30}$.
Main difference for the production of bio-molecules 
between Das et al., (2008a), Chakrabarti \& Chakrabarti (2000ab) 
and Majumdar et al., (2012, 2013) along with the present paper is 
the usage of rate coefficients during the formation of these molecules. 
To have an educated estimation for the production of bio-molecules (alanine, glycine \& adenine), 
in Das et al., (2008a) and Chakrabarti \& Chakrabarti (2000ab), rate coefficients were assumed to be 
$\sim 10^{-10}$ cm$^3$s$^{-1}$ 
whereas in Majumdar et al., (2013) and present paper, rate coefficients were taken 
from Majumdar et al., (2012).
Moreover, Since Das et al., (2008a) and Chakrabart \& Chakrabarti (2000ab) did not consider 
the extensive surface chemistry network, 
their computed gas phase abundances does not seems to be depleted whereas due to the consideration 
of a large surface chemistry network,
related chemical species of selected bio-molecules could be depleted on the grain surface which 
could affect their gas phase abundances.
Differences between the results of Majumdar et al., (2012, 2013) \&
the present paper are unsurprising because here, we are using totally different physical conditions
(elaborately mentioned in the earlier sections) in comparison to the Majumdar et al., (2012, 2013).

\section{Conclusion}
In this paper, we carry out numerical simulations to find out the abundances of different interstellar
molecules inside a collapsing and rotating interstellar cloud.
A well tested two dimensional hydrodynamics code has been used to obtain the physical properties
during the collapsing phase of a generic molecular cloud. Mechanisms which are responsible for the
interstellar heating and cooling are considered. The dynamic behaviour of the interstellar cloud
during the collapsing phase are used as the input parameter for the chemical code.
Major improvements over the chemical model considered in Das et al. (2008a)
are the inclusion of the grain chemistry self-consistently
and the inclusion of gas phase chemical network from the UMIST 2006 data base (Woodall et al., 2007),
Chakrabarti et al.,(2000ab), Majumdar et al., (2012). 
A detail comparison between our present work with our earlier work along with some other important 
modeling results are highlighted.

Temperature is an important physical parameters for a deciding the molecular complexity of a molecular
cloud. We carried out our simulation with both the constant temperature and varying the
temperature of the cloud. Results clearly shows strong variation in between these two consideration.

So far we have chosen only a single rotation parameter just to show its effect. With the increase of
rotational velocity, the centrifugal force would remove a large amount of matter in the
outer regions, especially along the axis of the cloud. These bi-polar flows would have significant effects
as they would carry away chemicals produced deep inside and distribute them at outer regions. Part of this
chemically enriched matter could be farther reprocessed as it is accreted again along with inflow. Furthermore,
a stronger centrifugal force would create shock waves changing the density and temperature distribution
dramatically which in turn would also modify chemical abundances. 
This aspect is being studied throughly and would be reported elsewhere.

\section{Acknowledgment}
SKC, LM \& SC are grateful to DST for the financial support through a 
project (Grant No. SR/S2/HEP-40/2008) and AD wants to thank ISRO 
respond project (Grant No. ISRO/RES/2/372/11-12). We would like to acknowledge 
D. Ryu from Chungnam National University, Daejeon, Korea for his 
suggestions in developing the TVD code with self-gravity.

\end{document}